\documentclass[letterpaper,12pt,onecolumn, nodraft]{IEEEtran}
\usepackage{graphicx,latexsym,epsfig,amssymb,amsmath,subfigure}
\usepackage{setspace}
\doublespace
\begin{document}

\title{Modeling Internet Security Investments  \\ \emph{The Case of Dealing with Information Uncertainty}}


\author{Ranjan Pal and Pan Hui\\ University of Southern California, Deutsch Telekom Laboratories\\Email: rpal@usc.edu, pan.hui@telekom.de}
\maketitle
\begin{abstract}
Modern distributed communication networks like the Internet and censorship-resistant networks  (also a part of the Internet) are characterized by nodes (users) interconnected with one another via communication links.  In this regard, the security of individual nodes depend not only on their own efforts, but also on the efforts and underlying connectivity structure of neighboring network nodes. By the term `effort', we imply the amount of investments made by a user in security mechanisms like antivirus softwares, firewalls, etc., to improve its security. However, often due to the large magnitude of such networks, it is not always possible for nodes to have complete effort and connectivity structure information about all their neighbor nodes. Added to this is the fact that in many applications, the Internet users are selfish and are not willing to co-operate with other users on sharing effort information. In this paper, we adopt a non-cooperative game-theoretic approach to analyze individual user security in a communication network by accounting for both, the partial information that a network node possess about its underlying neighborhood connectivity structure, as well as the presence of positive externalities arising from efforts exerted by neighboring nodes. We investigate the equilibrium behavior of nodes and show 1) the existence of symmetric Bayesian Nash equilibria of efforts and 2) better connected nodes choose lower efforts to exert but earn higher utilities with respect to security improvement \emph{irrespective} of the nature of node degree correlations amongst the neighboring nodes. Our results provide ways for Internet users to appropriately invest in security mechanisms under realistic environments of information uncertainty.

\end{abstract}
%
\IEEEpeerreviewmaketitle

\section{Introduction}  \label{sec-intro}
The Internet has become a fundamental and integral part of our daily lives. Billions of people are using the Internet for various types of applications that demand different levels of security. For example, commercial and government organizations run applications that require a high level of security, since security breaches would lead to large financial damage and loss of public reputation. Another example of a high security application in the Internet is maintaining user anonymity through a censorship-resistant network. On the other hand, an ordinary individual for instance generally uses a computing device for purposes that do not demand strict security requirements. However, all these applications are running on a network, that was built under assumptions, some of which are no longer valid for today's applications, e.g., that all users on the Internet can be trusted and that the computing devices connected to the Internet are static objects. Today, the Internet comprises of both good and malicious users. The malicious users perform illegal activities, are able to aspect many users in a short time period, and at the same time reduce their chances of being discovered. To overcome security related issues, Internet users invest in security mechanisms such as anti-virus solutions and firewalls. 

It is commonsense information that due to Internet connectivity, the security strength of an Internet user\footnote{An Internet user could be a single individual or an individual organization.} is dependent on the security strength of other users, especially neighboring users. Thus, from an individual user perspective, two important pieces of information are the amount of security investments of its neighbors in the network and the knowledge of the underlying connectivity structure of its neighbors, as they both drive optimal user investments. Unfortunately, due to the large magnitude of the Internet, its not feasible or practical to have exact information about the security investments and connectivity structure of all neighboring Internet users. In addition, most Internet users are selfish in nature and would not be inclined to share investment information with other Internet users. However, users do need to invest in security/defense mechanisms to protect themselves as much as possible. In this paper, we address the problem of optimal security investments when an individual user is uncertain about the underlying network connectivity structure of its neighbors , and accounts for the network externalities\footnote{An externality is a positive(negative) effect caused to a user not directly involved in an economic transaction, by other users involved in the transaction. For example, an Internet user investing in security mechanisms benefits all the nodes connected to it and thus creates a positive externality for its neighbors.} posed by them when they invest in security mechanisms. 

We consider models related to two general security scenarios as mentioned in \cite{hvar} when network externalities are present: 1) where the security strength of an individual user depends upon the sum security strength of itself and other individual nodes in the network under operation and 2) where the security strength of an individual user depends on the strength of the strongest node/s in the network. An example of scenario 1 is a peer-to-peer network where an attacker might want to slow down the transfer of a given piece of information, whose transfer speed might depend on the aggregate effort of all relevant nodes concerned. An example of scenario 2 is a censorship-resistant network, where a piece of information will remain available to a public domain as long as atleast one node serving that piece of information is unharmed. Another example of scenario 2 is the flow of traffic between two backbone nodes in the Internet. Modeling each path between two backbone nodes as a node, traffic will flow securely between the backbone as long as there is atleast one node that is unharmed by an attacker, i.e., there exists atleast one path between the backbone nodes. Likewise, there are other examples of  applications on the Internet that fit scenarios 1 and 2.  We emphasize here that there is another practical scenario as mentioned in \cite{hvar}, viz., one where the security strength of an individual user depends on the strength of the weakest node. This scenario is mainly an intra-organization scenario, where once a node in an organization is compromised due to a weak password or a security policy, its easy for an attacker to hack the whole system. However, the information of neighborhood topology structure within an organization may be known to the network users in certainty, but in this paper we focus on the case when users have uncertain information about the neighborhood topology structure of the network in operation.

\emph{Our Research Contributions}
\begin{enumerate}
\item We present a general model for analyzing individual user security in a non co-operative Internet environment.  In this regard, we study security games when 1) Internet users have incomplete information about the underlying neighboring network connectivity structure and 2) Internet users account for the positive externalities posed by the investments of neighboring Internet users. Our model extends work proposed in \cite{gccr1}\cite{gccr}  in terms of capturing network information uncertainty. (See Section III.) \\
\item We formulate our investment problem as a Bayesian game of incomplete information and show the existence of a symmetric Nash equilibrium of user investments. The equilibrium results show that under incomplete neighboring network topology information, better connected users choose lower efforts to exert and earn higher payoffs, \emph{irrespective} of the nature of node degree correlations amongst neighboring nodes. \footnote{In a network such as the Internet, there exists a correlation between the node degrees \cite{newman2}. In this paper we explicitly model the degree correlations.}. (See Section IV.)  
\end{enumerate}
 
\section{Related Work}  \label{related-work} 
There have been very few works related to security investments in the Internet. The authors \cite{leb}\cite{pg} in their works have analyzed self-protection investments in Internet security under the presence of cyber-insurance, which is a form of a third-party risk transfer. Under the assumption of users having complete network topology information, the works show 1) cyber-insurance incentivizes users to invest in self-protection, 2) cyber-insurance entails optimal user investments both in insurance and in self-protection, and 3) co-operation amongst Internet users result in higher user self-protection investments when compared to the case when users do not co-operate. However, attractive though the concept may seem, cyber-insurance may not be a market reality due to factors such as inter-dependent security, correlated risks, and information asymmetry between the insurer and the insured \cite{rabohme}\cite{ssfw}. In addition, it is also infeasible for Internet users to have complete network topology information.

For non cyber-insurance environments, in a recent series of works \cite{jaw}\cite{oom}, the authors show that Internet users invest sub-optimally in security under selfish environments when compared to the case when user co-operation is allowed. They account for externalities but base their results by assuming users having complete network topology information. However, as we have discussed previously, in a large network such as the Internet, having complete network topology information is infeasible. In addition, all the mentioned related works do not model the well-known security games mentioned in \cite{hvar}, that are in general played by attackers and defenders (non malicious Internet users) when externalities are present in a network. In this regard, the works in \cite{zg}\cite{gccr1}\cite{gccr}  tackle the problem of optimal security investments and model the cited security games mentioned in \cite{hvar} but do not account for any uncertainty of information that a user has regarding the underlying network topology. In this paper, we advance previous research in security investments and model both, externalities, as well as users having uncertainty of information regarding the connectivity structure of neighboring nodes. However, unlike \cite{gccr1}\cite{gccr}, we do not model self-insurance, and only focus on self-protection without any cyber-insurance. 

\section{Modeling Network Security Investment Games}  \label{sec-model}
In this section, we propose a general model for analyzing network security investments using a game-theoretic approach. First, we model the user interaction network in the Internet. Second, we describe the utility/payoff function of the Internet users as a function of their strategies/actions, which are nothing but the security investments of a user. Finally, we explain the information structure of Internet users with respect to the underlying connectivity structure of their neighbors, and highlight the game of investments that results from the information structure. 

\subsection{Network Structure} \label{sec-ns}
We consider a set $N = \{1, ......, n\}$ of $n$ Internet users and a connectivity matrix $G = (V,E)$ of users, where $v_{ij} = 1$ if the utility of user $i$ is affected by the security investment of user $j$, $i$ being not equal to $j$, and 0 otherwise. Let $N_{i}(v) = \{j|v_{ij} = 1\}$ denote the set of all the one hop neighbors of $i$, where $v\,\epsilon\,\{0,1\}^{n \times n}$. We also consider the $k$-hop neighbors of node $i$ and denote the set by $N_{i}^{k}(v)$. This set consists of all the nodes that are within $k$-hops of node $i$, where $k\ge1$. Inductively,  we have the following relationships between $N_{i}^{k}(v)$ and $N_{i}(v)$:
\begin{equation}
N_{i}^{1}(v) = N_{i}(v).
\end{equation}
\begin{equation}
N_{i}^{k}(v) = N_{i}^{k - 1}(v) \cup (\cup_{j\,\epsilon\,N_{i}^{k - 1}(v)}N_{j}(v)).
\end{equation}
We represent the degree of a node $i$ by $d_{i}$, where $d_{i}$ equals $|N_{i}(v)|$. In this paper, we assume that each user has perfect knowledge about its own degree but does not have complete information about the degrees of its neighbors. (More on degree information structure in Section \ref{sec-is}.)

\subsection{User Strategies and Payoffs} \label{sec-payoff}
In this paper we consider two types of non co-operative security investment games concerning the case when users have incomplete information on the topology of the network under operation: (1) the users are selfish and invest to maximize their own utilities, but the security strength of an individual user depends on the sum of security investments of itself and its neighboring individual nodes and 2) the users are selfish and invest to maximize their own utilities, but the security strength of the whole network depends on the security investments of the most robust node/s amongst its neighbors. The latter type of game is often termed as a `best-shot' game. In both these types of games, each user is a player and its strategy is the amount of security investment it makes. We assume here that the strategy/action of each user $i$ is $x_{i}$ and lies in the \emph{compact}\footnote{In mathematical analysis, a compact set is one which is closed and bounded.} set $[0,1]$. We also assume that the utility/payoff to each user $i$ is $U_{i}$ and is a function of the security investments made by itself and its one hop neighbors. Thus $U_{i} = U_{i}(x_{i}, \overrightarrow x_{N_{i}(v)})$, where $\overrightarrow x_{N_{i}(v)}$ is the vector of security investments of the one hop neighbors of user $i$. From the structure of user utility functions, we observe that two players having the same degree will have the same utility function.  We also model the concept of a positive externality as it forms an integral part of the analysis in Section IV. A positive externality to a user from its one hop neighbors results when they invest in security, thereby improving the individual security strength of the user. We represent the concept mathematically in the following manner: we say that a payoff function exhibits positive externalities if for each $U_{i}$ and for all $\overrightarrow x \ge \overrightarrow x', U_{i}(x_{i}, \overrightarrow x) \ge U_{i}(x_{i}, \overrightarrow x')$, where $\overrightarrow x$ and $\overrightarrow x'$ are the vectors of security investments of one hop neighbors of user $i$.

In scenarios where the security strength of a user $i$ depends on the sum of investments of itself and other neighboring users, we mathematically formulate $i$'s utility/payoff function as follows:
\begin{equation}
U_{i}(x_{1},......., x_{d_{i}}) = f\left(x_{i} + \lambda\sum_{j = 1}^{d_{i}}x_{j}\right) - c(x_{i}),
\end{equation}
where $f(\cdot)$ is a non-decreasing function, $c(\dot)$ is the cost incurred by user $i$ for putting in own effort in order to make its system more robust, and $\lambda$ is a scalar quantity which determines the magnitude of the positive externality experienced by user $i$ due to the security investments made by its one hop neighbors.  
  
The situation when the security strength of a user depends on the investments made by the strongest neighbor/s can be modeled as a special case of the situation when a user security strength depends on the sum of the security investments of its neighbors. We first note that from user $i$'s perspective, the former situation implies that as long as there is a neighboring node/s that is secure, user $i$ is safe. In Section \ref{sec-intro} we have already cited censorship resistant networks and Internet backbone networks to be examples of networks where the former situation might arise leading to a best-shot game. We had also given an example of how the best-shot scenarios arising in these networks can be modeled as a graph to reflect the `user-neighbor' concept.

Once we have modeled a best-shot scenario as a graph, we fix the strategy space of individual users to $\{0,1\}$ and make $f(0) = 0$ and $f(y)$ = 1 for all $y \ge 1$. A binary strategy space of $\{0,1\}$ implies that each user decides either to invest or not to invest. If a user or any of its neighbors invest, the former is safe, else it is not. We observe that the `sum of investments' game gets converted to a best-shot game. In this case user $i$'s payoff follows the following equation:
\begin{equation}
U_{i}(x_{i}, (\overrightarrow x,0)) = U_{i}(x_{i}, \overrightarrow x),\, \forall (x_{i}, \overrightarrow x)\,\epsilon\,[0,1]^{d_{i} + 1}. 
\end{equation}
Equation (4) implies that adding a link to a neighbor who invests zero amount in security mechanisms is equivalent to not having the neighbor.  

\subsection{Information Structure} \label{sec-is}
In this paper we assume that each Internet user (player) knows its own degree but does not have perfect information regarding the degree of its neighbors. It has already been shown by Newman in \cite{newman2} that nodes in an Internet like network exhibit degree correlations\footnote{Newman show through empirical studies that technological and Internet networks exhibit negative degree correlation whereas social networks exhibit positive degree correlation.} In this regard, we account for the degree correlations between the neighboring nodes of a user $i$ in our model, i.e., when a user decides on its strategy, it accounts for the amount of information it has on the degree of its neighbors. Information on degree correlations is important as it guides a user to making better security investments when compared to the situation when it has no information about the correlations.   

Let the degrees of the neighbors of user $i$ be the vector $\overrightarrow d_{N_{i}(v)}$, whose dimension is $ d_{i}$. We assume that user $i$ does not know the vector $\overrightarrow d_{N_{i}(v)}$ but has information regarding its probability distribution, i.e., it knows the value of $P(\overrightarrow d_{N_{i}(v)}|d_{i})$. We assume that each player in the network under consideration has symmetrical beliefs about the degree of its neighbors. Thus, arises a family of conditional distributions, \textbf{C} $\equiv \{[P(\overrightarrow d|d)]_{\overrightarrow d\,\epsilon\,\mathbb{N}^{d}}\}_{d\,\epsilon\,\mathbb{N}}$, where $\overrightarrow d$ is a vector of degrees of the neighbors of a node and $d$ is the degree of a given node.

We model the strategic interactions between the players of the network as a \emph{Bayesian game of incomplete information}. The type space of the Bayesian game is the user knowledge on the potential degrees of its neighboring players. The strategy for each player is its security investment conditioned on the knowledge of the degree of their neighbors, and the payoff function for each player is as defined in Section \ref{sec-payoff}, which depends on whether the game is a sum of investments game or a best-shot game. Assuming that $S$ is the set of possible investments a user could make, the strategy for player $i$ is a mapping $\gamma_{i}:\,\{0,1,....., n - 1\}\rightarrow \Omega(S)$, where $ \Omega(S)$ is the set of distribution functions on $S$. 
  
We note that for a player, its conditional distributions concerning the neighbors' degrees can vary with its own degree. According to our model, players may have different number of neighbors, and the degrees of the neighbors are correlated with each other. Thus, the dimension of the vector of degrees of its neighbors may vary from player to player. In order to address correlation amongst vectors of different dimensions, we adopt the technique of `association' from the domain of statistics \cite{epw}. Association is used to track the correlation patterns of groups of random variables, given the complicated interdependencies that might be present between them. A positive association indicates that higher levels of one variable (in this case a player's degree) implies higher levels of all other variables (in this case a player's neighbors' degrees).  

Given a player $i$ with degree $d_{i}$, enumerate the degrees of $i$'s neighbors as $\overrightarrow d_{N_{i}(v)} = (d_{1},........,d_{d_{i}})$. Now consider a function $F: \{0,1,......, n- 1\}^{m} \rightarrow \mathbb{R}$, where $m \le d_{i}$. Let 
\begin{equation}
E_{P(\cdot|d_{i})}[F] = \sum_{\overrightarrow d_{N_{i}(v)}} P(\overrightarrow d_{N_{i}(v)}|d_{i})F(d_{1},......,d_{m}).
\end{equation}
In Equation (5) we fix a subset $m\le d_{i}$ of user $i$'s neighbors, and then take the expectation of $F$ operating on their degrees. We say that the family of distributions \textbf{C} exhibits positive association if, for all $k' > k$, and any non-decreasing $F:\{0,1,.........,n -1\}^{k} \rightarrow \mathbb{R}$, we have 
\begin{equation}
E_{P(\cdot|k')}[F] \ge E_{P(\cdot|k)}[F],
\end{equation}
and \textbf{C} exhibits negative association if  
\begin{equation}
E_{P(\cdot|k')}[F] < E_{P(\cdot|k)}[F],
\end{equation}
for  all $k' > k$, and any non-decreasing $F:\{0,1,.........,n -1\}^{k} \rightarrow \mathbb{R}$.
\section{Game Analysis}
In this section, we analyze the\emph{symmetric} Bayesian game of incomplete information played between the users of the network under operation. In any symmetric game, the player payoffs for playing a particular strategy depend only on the strategies of other players and not on who is playing the strategies. We investigate the existence, uniqueness, and monotonicity of our game equilibria. In studying monotonicity of equilibria, we investigate the changes in the best response investment magnitude of a user when other users in the network increase/decrease their best response investment amounts. We also investigate the effect of the increase/decrease in user degrees on the equilibria of the game. We initially give a mathematical definition of our Bayesian game and follow it up with the analysis of game equilibria. 

\subsection{Game Definition}
Consider a player (Internet user) $i$ having degree $d_{i}$ in a \emph{sum-of-investments game} or a \emph{best-shot game}. Each player chooses a security investment amount from the set $S$ as its strategy, where $S$ is as defined in Section \ref{sec-is}. Let $d\rho_{-i}(\overrightarrow \gamma, d_{i})$ be the probability density over $x_{N_{i}(v)}\,\epsilon\,S^{d_{i}}$ induced by the beliefs $P(\cdot|d_{i})$ held by player $i$ over the degrees of its neighbors, combined with the strategies played via $\overrightarrow \gamma$, the vector of strategies of other users in the network. Let 
\begin{equation}
EU_{i}(x_{i}, \overrightarrow \gamma, d_{i}) = \int_{x_{N_{i}(v)}\,\epsilon\,S^{d_{i}}} U_{i}(x_{i}, x_{N_{i}(v)})d\rho_{-i}(\overrightarrow \gamma, d_{i}),
\end{equation}
where $EU_{i}(x_{i}, \gamma, d_{i})$ is the expected utility/payoff of player $i$ with degree $d_{i}$ and investment $x_{i}$ when other players choose strategy $\overrightarrow \gamma$. The \emph{Bayesian Nash equilibrium} of the game is a strategy vector that \emph{maximizes} the expected utility of each player in the network \cite{ft}\cite{or}.  In regard to individual user expected payoff functions, we now define the concepts of degree complementarity and degree substitutability that will be of importance in the analysis of the \emph{monotonicity} of game equilibria. 

For a given player $i$, we say that its expected utility function exhibits degree complementarity if 
\begin{equation}
EU_{i}(x_{i}, \overrightarrow \gamma, d_{i}) - EU_{i}(x'_{i}, \overrightarrow \gamma, d_{i}) \ge EU_{i}(x_{i}, \overrightarrow \gamma, d'_{i}) - EU_{i}(x'_{i}, \overrightarrow \gamma, d'_{i}),
\end{equation}
where $x_{i} > x'_{i}$, $d_{i} > d'_{i}$, and $\overrightarrow \gamma$ is non-decreasing. Similarly for a given player $i$, we say that its expected utility function exhibits degree substitutability if 
\begin{equation}
EU_{i}(x_{i}, \overrightarrow \gamma, d_{i}) - EU_{i}(x'_{i}, \overrightarrow \gamma, d_{i}) \le EU_{i}(x_{i}, \overrightarrow \gamma, d'_{i}) - EU_{i}(x'_{i}, \overrightarrow \gamma, d'_{i}),
\end{equation}
where $x_{i} > x'_{i}$, $d_{i} > d'_{i}$, and $\overrightarrow \gamma$ is non-increasing. We have the following lemma and ensuring the conditions under which the expected utility of a player exhibits degree complementarity. \\  
\textbf{Lemma 1.}  \emph{Given the conditions that} (1) $U_{i}(x_{i}, (\overrightarrow x,0)) = U_{i}(x_{i}, \overrightarrow x),\, \forall (x_{i}, \overrightarrow x)\,\epsilon\,[0,1]^{d_{i} + 1}$, \emph{for each player} $i$, (2) \emph{the} $U_{i}(\cdot)$'s \emph{for each player} $i$ \emph{exhibit strategic complements\footnote{$U_{i}(\cdot)$ is said to exhibit strategic complements \cite{hv} if for all $d_{i}$, $x_{i} > x'_{i}$ and $\overrightarrow x \ge \overrightarrow x'$ implies $U_{i}(x_{i}, \overrightarrow x) - U_{i}(x'_{i}, \overrightarrow x) \ge U_{i}(x_{i}, \overrightarrow x')  - U_{i}(x'_{i}, \overrightarrow x')$. Analogously, $U_{i}(\cdot)$ is said to exhibit strategic substitutes \cite{hv} if for all $d_{i}$, $x_{i} > x'_{i}$ and $\overrightarrow x \ge \overrightarrow x'$ implies $U_{i}(x_{i}, \overrightarrow x) - U_{i}(x'_{i}, \overrightarrow x) \le U_{i}(x_{i}, \overrightarrow x')  - U_{i}(x'_{i}, \overrightarrow x')$.}, and} (3) \emph{the family of conditional distributions} \textbf{C} \emph{is positively associated, then} $EU_{i}$'s \emph{for each player} $i$ \emph{exhibits degree complements.}  \\ 

We omit the proof due to lack of space. We emphasize here that the proof structure also establishes that given (1) $U_{i}(x_{i}, (\overrightarrow x,0)) = U_{i}(x_{i}, \overrightarrow x),\, \forall (x_{i}, \overrightarrow x)\,\epsilon\,[0,1]^{d_{i} + 1}$, for each player $i$, (2) the $U_{i}(\cdot)$'s for each player $i$ exhibit strategic substitutes, and (3) the family of conditional distributions \textbf{C} is negatively associated, $EU_{i}$'s for each player $i$ exhibits degree substitutes.
\subsection{Game Equilibria Results}
In this section we state the results related to equilibria of our proposed Bayesian game. Given a symmetric environment; i.e., players participate in a symmetric Bayesian game of security investments, we prefer to analyze \emph{symmetric equilibria}\footnote{A symmetric equilibrium is one where each player in the game plays the same strategy.}, as asymmetric behavior seems relatively unintuitive, and difficult to explain in a one-shot interaction \cite{kreps}. We omit the proofs of the results due to lack of space. \\ 
\textbf{Lemma 2.} \emph{There exists a symmetric equilibrium in our proposed security investment game. If the expected utility function of users exhibit degree complementarity, the equilibrium is non-decreasing, whereas for user utility functions exhibiting degree substitutes, the equilibrium is non-increasing.}\\ \\
\textbf{Lemma 3.} \emph{Given the conditions that} (1) $U_{i}(x_{i}, (\overrightarrow x,0)) = U_{i}(x_{i}, \overrightarrow x),\, \forall (x_{i}, \overrightarrow x)\,\epsilon\,[0,1]^{d_{i} + 1}$, \emph{for each player} $i$ and (2) \emph{degrees of neighboring nodes of users are independent, then strategic substitutes (compliments) of user utility functions result in every symmetric equilibrium of our proposed Bayesian game being monotone increasing (decreasing).}  \\ \\
\textbf{Lemma 4.}  \emph{Suppose} $U_{i}(x_{i}, (\overrightarrow x,0)) = U_{i}(x_{i}, \overrightarrow x),\, \forall (x_{i}, \overrightarrow x)\,\epsilon\,[0,1]^{d_{i} + 1}$, \emph{for each player} $i$. \emph{If} \textbf{C} \emph{is positively associated, then in every non-decreasing symmetric equilibrium of our proposed Bayesian game, the expected utilities of players are non-decreasing in degree. If} \textbf{C} \emph{is negatively associated, then in every non-increasing symmetric equilibrium of our proposed Bayesian game, the expected utilities of players are non-decreasing in degree.} \\ \\
\emph{Lemma Comments:} From a user (player) perspective, Lemma 2 states that when user expected utilities exhibit degree complementarity, a monotonic increase in the equilibrium security investments of all other users results in an increase in the player's equilibrium investments, for at least one equilibrium. Thus, the degree complementarity property of user expected utilities prevents \emph{free-riding} behavior of users for at least one equilibrium. On the other hand, the degree substitutes property ensures that for at least one equilibrium, users are not incentivized to increase their security investments when others in the network do not, in turn providing no additional benefit to other network users by investing relatively more. Lemma 3 states the conditions under which all symmetric equilibria are monotone, and gives an insight on the topology of the network that could result in all symmetric equilibria being monotone. Lemma 4 provides the relation between network degrees of users and their equilibrium payoffs, and identifies the conditions under which payoffs increase/decrease with network degree. The relationships state the contexts in which network connections are advantageous and disadvantageous with respect to equilibrium payoffs. 
\section{Conclusion}
In this paper we proposed a security investment model for the Internet in which Internet users account for the positive externality posed to them by other Internet users and make security investments under situations when they do not have complete information about the underlying connecting topology of its neighbors. Our model is based on a game-theoretic approach and we showed 1) the existence of symmetric Bayesian Nash equilibria of efforts and 2) better connected nodes choose lower efforts to exert but earn higher utilities with respect to security improvement \emph{irrespective} of the nature of node degree correlations amongst the neighboring nodes. Our results provided ways for Internet users to appropriately invest in security mechanisms under realistic environments of information uncertainty. Our results also clarified how the basic strategic features of the game - as manifest in the complements and substitutes property - combine with different patterns of degree association to shape network behavior and user payoffs. As a part of future work, we plan to incorporate the network concepts of centrality and clustering in our model in addition to degree distributions. We also plan to investigate security investments under an asymmetric environment, i.e., a game environment in which user payoffs depend not only on the strategy of other users but also on the identity of the users.

\newpage
\bibliography{alluvion}
\bibliographystyle{plain}


\end{document}